\documentclass[12pt]{iopart}

\usepackage{graphicx}

\begin{document}

\title[]{Fermion localization on asymmetric two-field thick branes}

\author{Zhen-Hua Zhao,
        Yu-Xiao Liu\footnote{Corresponding author.} and
        Hai-Tao Li}

\address{
    Institute of Theoretical Physics,
    Lanzhou University,
    Lanzhou 730000, P. R. China}

\ead{zhaozhh09@lzu.cn, liuyx@lzu.edu.cn and liht07@lzu.cn}
\begin{abstract}
In this paper we investigate the localization of fermions
on asymmetric thick branes generated by two scalars $\phi$ and
$\chi$. In order to trap fermions on the asymmetric branes with
kink-like warp factors, the couplings with the background scalars
$\eta\bar{\Psi}F(\chi,\phi)\Psi$ are introduced, where $F(\chi,\phi)$
is a function of $\phi$ and $\chi$. We find that the coupling
$\eta\bar{\Psi}\chi\phi\Psi$ do not support the localization of
4-dimensional fermions on the branes. While, for the case
$\eta\bar{\Psi}\chi\Psi+\eta'\bar{\Psi}\phi\Psi$, which is the
kink-fermion coupling corresponding to one-scalar-generated brane
scenarios, the zero mode of left-handed fermions could be trapped on
the branes under some conditions.

\end{abstract}

\pacs{11.10.Kk,04.50.-h,11.27.+d}
\maketitle

\section{Introduction}

During the last two decades, the braneworld scenarios have received
considerable attention. The idea of embedding our universe in a
higher dimensional space-time provides new insights for solving
long-standing problems such as the gauge hierarchy problem
\cite{ADD,RS} and cosmological constant problem \cite{Rubakov1983}.
Recently the thick brane scenario has obtained an increasing
attention \cite{DeWolfe200062,Csaki2000581,Gremm2000478,Campos200288,
ThickBranes,ThickBraneWeyl,Liu0907.1952}, since in more realistic
models the thickness of the brane should be taken into account. One
can refer to a recent review article on the subject of the thick
brane solutions~\cite{Dzhunushaliev:09a}. Two real scalar fields
coupled with gravity in 5-dimensional warped space-time as a thick
brane realization was investigated by Bazeia and Gomes in Ref.
\cite{Bazeia200405}, where the brane was called a Bloch brane. The
Bloch brane is an extension of the known Bloch wall~\cite{Blochwall}
in the context of braneworlds.

\begin{figure}
\begin{center}
\includegraphics{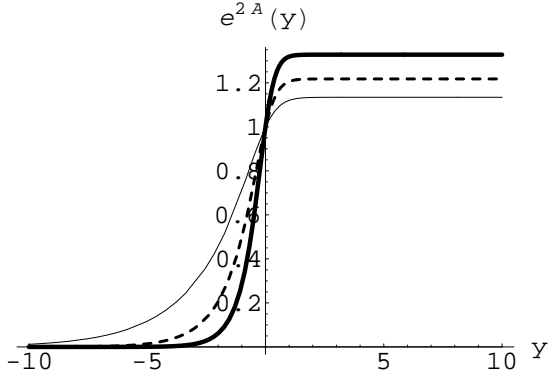}
\end{center}
\caption{ \label{fig:A_y} The shape of the warp factor $e^{2A(y)}$.
The parameters are set to $\mu=1$, $a=1.2$ (thick line), $a=1$
(dashed line) and $a=0.8$ (thin line).}
\end{figure}

Two kinds of analytic solutions of asymmetric two-field branes were
found in Ref.~\cite{SouzaDutra200878}. In Fig.~\ref{fig:A_y}, one of
the warp factors $e^{2A}$ for the asymmetric two-field branes is
plotted. From Fig.~{\ref{fig:A_y}}, we can see that the warp factor
tends to zero and a constant when $y$ goes to $-\infty$ and $\infty$,
respectively. However, in most of other brane solutions, the warp
factors decrease to zero when $y$ approaches to $\pm\infty$ or the
boundaries of the extra dimension.

In a braneworld theory, a very interesting and important problem is
the localization of fermions. We know that Localization of fermions
on thick branes requires additional interactions besides gravity: a
scalar-fermion coupling has to be included. For some thick branes,
with the scalar-fermion coupling, there may exit single bound state
(the massless mode) \cite{LocalizedFermion,Liu2008D78,Almeida200979,LiuJHEP2009},
while for some other ones, mass gap and finite discrete Kaluza-Klein
(KK) modes appear in the KK spectrum of fermions
\cite{Parameswaran0608074,GeorgePRD2007,Liu200808,Liu20090902,KodamaPRD2009,Ringeval200265,Liu0907.4424}.
In Ref. \cite{Almeida200979}, the simplest Yukawa coupling
$\bar\Psi\phi\chi\Psi$ was introduced for the two-field brane model,
and one single bound state of left-handed fermions was obtained.

As the asymmetric two-field branes found in \cite{SouzaDutra200878}
are very different from symmetric thick branes, an interesting
problem is to investigate how spin-1/2 fermions can be localized on
the branes. We will show that the coupling
$\eta\bar{\Psi}\chi\phi\Psi$ considered in the literature for
two-scalar brane models do not support the localization of
4-dimensional massless fermions on the branes. So, other couplings
should be considered in order for massless chiral fermions to
localize on the branes. We find that for the coupling
$\eta\bar{\Psi}\chi\Psi+\eta'\bar{\Psi}\phi\Psi$, the kink-fermion
coupling corresponding to one-scalar-generated brane scenarios, the
zero mode of left-handed fermions can be trapped on the branes
providing that the coupling constants $\eta$ and $\eta'$ satisfy some
conditions.

This paper is organized as follows: We first review the asymmetric
two-field brane model. Then we study the localization and spectra of
fermions on the asymmetric two-field branes by presenting the
mass-independent potentials of the corresponding Schr\"{o}dinger
problem of fermionic KK modes. We consider several different types of
scalar-fermion interaction. Finally, the conclusion is given in the
last section.


\section{Review of the asymmetric two-field thick brane model}\label{section:review}

Bloch branes are constructed by two interacting real scalar fields
$\phi$ and $\chi$  with a scalar potential $V(\chi,\phi)$
\cite{Bazeia200405}. The corresponding action for the system is
\begin{eqnarray}
 S=\int d^4xdy\sqrt{-g}\bigg[\frac{1}{4}R
 -\frac{1}{2}g^{MN}(\partial_{M}\phi~\partial_{N}\phi 
 +\partial_{M}\chi\partial_{N}\chi)
 -V(\chi,\phi)\bigg],\label{action}
\end{eqnarray}
where $y$ is the coordinate of the extra dimension, $g\equiv
\textrm{det}(g_{MN})$, and the metric is assumed as
\begin{eqnarray}
 ds^2=g_{MN}dx^Mdx^N=e^{2A}\eta_{\mu\nu}dx^{\mu}dx^{\nu}
 +dy^2\label{metric1}
\end{eqnarray}
with $e^{2A}$ the warp factor. A usual hypothesis is that $A$, $\phi$
and $\chi$ are only the functions of $y$. From the action
(\ref{action}) we can obtain the Einstein equations and the equations
of motion of scalar fields
\cite{DeWolfe200062,Bazeia200405,SouzaDutra200878}:
\begin{eqnarray}
\begin{array}{l}
  \chi''+4A'\chi'=V_{\chi}(\chi,\phi),\\
 \phi''+4A'\phi'=V_{\phi}(\chi,\phi),\\
 A''=-\frac{2}{3}(\phi'^2+\chi'^2),\\
 A'^2=\frac{1}{6}(\phi'^2+\chi'^2)-\frac{1}{3}V(\chi,\phi),
\end{array}\label{eq:2order}
\end{eqnarray}
where prime stands for the derivative with respect to $y$, and
$V_{\phi}\equiv\partial V/\partial \phi$. By introducing the
superpotential
\begin{eqnarray}
 W(\chi,\phi)= \phi\left[\lambda\left(\frac{\phi^2}{3}
                -a^2 \right)+\mu\chi^2 \right],
\end{eqnarray}
the scalar potential \cite{SouzaDutra200878}
\begin{eqnarray}
 V(\chi,\phi)= \frac{1}{2}\left[
    \left(W_{\chi}(\chi,\phi)\right)^2
   +\left(W_{\phi}(\chi,\phi)\right)^2\right]
 -\frac{4}{3}W(\chi,\phi)^2 \nonumber
\end{eqnarray}
leads to another set of first-order differential equations to share
the solutions with the second-order equations (\ref{eq:2order}),
which are given by \cite{SouzaDutra200878}
\begin{eqnarray}
 \chi'=W_{\chi}(\chi,\phi),~~
 \phi'=W_{\phi}(\chi,\phi),~~
 A'=-\frac{2 }{3}W(\chi,\phi).
\end{eqnarray}
The above equations have four sets of asymmetric solutions when we
take $\lambda=\mu$ and $\lambda=4\mu$ \cite{SouzaDutra200878}.

For $\lambda=\mu$, the classical solutions for $\chi(y)$, $\phi(y)$
and $A(y)$ can be shown to be
\begin{eqnarray}
 \chi(y)&=&\frac{a}{2}\big[1\pm \tanh(a{\mu}y)\big],\\
 \phi(y)&=&\frac{a}{2}\big[\pm1-\tanh(a{\mu}y)\big],\label{eq:phi1}
\end{eqnarray}
and
\begin{eqnarray}
 A(y)=\frac{1}{18} a^2 \bigg[ 4 \ln  (1\pm\tanh(a{\mu}y))
 -\tanh^2(a{\mu}y) \bigg].\label{eq:A1}
\end{eqnarray}

For $\lambda=4\mu$, the solutions are
\begin{eqnarray}
 \chi(y)&=&\sqrt{2} ~a \frac{\cosh(a{\mu}y)\pm \sinh(a{\mu}y)}
 {\sqrt{\cosh(2a{\mu}y)}},\\
\phi(y)&=&\frac{a}{2}[\pm1 -\tanh(2a{\mu}y)],\label{eq:phi2}
\end{eqnarray}
and
\begin{eqnarray}
A(y)&=&\frac{ a^2}{36} \bigg\{\pm\bigg[32 a{\mu}y
             -6 \tanh(2a{\mu}y)\bigg] \nonumber\\
 &&+\tanh^2(2a{\mu}y)
 -16\ln\cosh(2a{\mu}y)\bigg\}.\label{eq:A2}
\end{eqnarray}

\section{Localization of fermions on the asymmetric two-field branes}
\label{section:Localization}

Let us consider the action of a 5-dimensional massless spin 1/2
fermion coupled to gravity and the background scalar fields:
\begin{eqnarray}
 S_{1/2}=\int d^5x\sqrt{-g}\left[\bar\Psi\Gamma^MD_M\Psi-\eta\bar\Psi
 F(\chi,\phi)\Psi\right],
 \end{eqnarray}
where $\eta$ is the coupling constant between fermions and scalar
fields. Here, the backreaction of the spinor field to the brane
solution has been neglected. In order to get mass-independent
potentials for KK modes of fermions, we will follow Ref.~\cite{RS}
and change the metric (\ref{metric1}) to a conformally flat one
\begin{eqnarray}
ds^2=e^{2A}(\eta_{\mu\nu}dx^{\mu}dx^{\nu}+dz^2)\label{metric2}
\end{eqnarray}
by performing the coordinate transformation
\begin{eqnarray}
dz=e^{-A(y)}dy.\label{eq:relation}
\end{eqnarray}
Using the conformal metric (\ref{metric2}) and the general gamma
matrices $\Gamma^M=(e^{-A}\gamma^{\mu},e^{-A}\gamma^{5})$, the
equation of motion for fermions is read as
\begin{eqnarray}
 \bigg[\gamma^{\mu}\partial_{\mu}+\gamma^5(\partial_z+2\partial_zA)-\eta
e^{A}F(\chi,\phi)\bigg]\Psi=0.
\end{eqnarray}
Now we will using the general chiral decomposition
\begin{eqnarray}
 \Psi(x,z)=\sum_n \bigg(\psi_{Ln}(x){f}_{Ln}(z)
 +\psi_{Rn}(x)f_{Rn}(z)\bigg)
\end{eqnarray}
to search for the solutions of right-handed and left-handed fermion
fields. Here $\psi_{Ln}=-\gamma^5\psi_{Ln}$ and
$\psi_{Rn}=\gamma^5\psi_{Rn}$, and they satisfy the 4-dimensional
massive Dirac equations:
$\gamma^{\mu}\partial_{\mu}\psi_{Ln}=m_{n}\psi_{Rn}$ and
$\gamma^{\mu}\partial_{\mu}\psi_{Rn}=m_{n}\psi_{Ln}$. Then we obtain
the equations for $f_{Ln}$ and $f_{Rn}$
\begin{eqnarray}
\{{\partial_{z}+2\partial_z A+\eta e^A
F(\chi,\phi)}\}f_{Ln}(z)&=&~~\,\;m_nf_{Rn}(z),\label{eq:zeroMode}\\
\{{\partial_{z}+2\partial_z A-\eta e^A
F(\chi,\phi)}\}f_{Rn}(z)&=&-m_nf_{Ln}(z),~~~~~
\end{eqnarray}
with the following orthonormality conditions:
\begin{eqnarray}
 \int_{-\infty}^{\infty}e^{4A}f_{Lm}f_{Ln}dz
    =\int_{-\infty}^{\infty}e^{4A}f_{Rm}f_{Rn}dz
    =\delta_{mn},\;
 \int_{-\infty}^{\infty}e^{4A}f_{Lm}f_{Rn}dz
    =0. \label{orthonormalityConditions}
\end{eqnarray}
Defining $f_{L}=\tilde{f}_{L}e^{-2A}$ and
$f_{R}=\tilde{f}_{R}e^{-2A}$, we obtain the Schr\"{o}dinger-like
equations for the new functions $\tilde{f}_{L}$ and $\tilde{f}_{R}$
\cite{Liu2008D78}:
\begin{eqnarray}
  \begin{array}{l}
  \big[-\partial_z^2+V_L(z)\big]\tilde{f}_{Ln}
      = m^2_n\tilde{f}_{Ln},\\
  \big[-\partial_z^2+V_R(z)\big]\tilde{f}_{Rn}
      = m^2_n\tilde{f}_{Rn},\\
  \end{array}  \label{eq:motionF}
\end{eqnarray}
where the effective potentials are given by
\begin{eqnarray}
\begin{array}{l}
  V_{L}(z)= \eta^2 e^{2A}F^2(\chi,\phi)-\eta\partial_z[e^A F(\chi,\phi)],\label{eq:Va}\\
  V_{R}(z)= \eta^2 e^{2A}F^2(\chi,\phi)+\eta\partial_z[e^A F(\chi,\phi)]. \label{eq:Vb}\\
\end{array}   \label{eq:V}
\end{eqnarray}

Unfortunately, because of the complexity of $A(y)$ in (\ref{eq:A1})
and (\ref{eq:A2}), we can not use the relation (\ref{eq:relation}) to
get an analytic form of $z(y)$ or its inverse $y(z)$. So we have to
use numerical method to produce the pair of points $(z, y)$ with a
constant step in $z$. This leads to the determination of $V_L(z)$ and
$V_R(z)$.  With the potentials in $z$ coordinate, we can solve
Eqs.~(\ref{eq:motionF}) numerically. With the expressions
${\partial}_z A =e^{A(y)}{\partial}_y A $ and ${\partial}_z
F=e^{A(y)} {\partial}_y F$, we can rewrite the potentials
(\ref{eq:V}) as the functions of $y$:
\begin{eqnarray}
\begin{array}{l}
     V_{L}(z(y))=\eta e^{2A}\big[\eta F^{2}-
            \partial_{y}F-F\partial_{y}A(y)\big], \\ 
     V_{R}(z(y))=V_{L}(z(y))|_{\eta\rightarrow-\eta}.\\ 
\end{array}\label{Vy}
\end{eqnarray}

The asymmetric two-field branes have four sets of solutions: two for
$\lambda=\mu$ and two for $\lambda=4\mu$. Because these solutions are
very similar, here we only consider the following one corresponding
to $\lambda=\mu$:
\begin{eqnarray}
 \chi(y)&=&\frac{a}{2}\big[1+ \tanh(\mu ay)\big],\label{eq:chi}\\
 \phi(y)&=&\frac{a}{2}\big[1-\tanh(\mu a y)\big],\label{eq:phi}\\
 A(y)&=&\frac{a^2}{18}  \big[ 4 \ln  \big(1+\tanh(a{\mu}y)\big)
 -\tanh^2(a{\mu}y) \big].~~~\label{eq:A(y)}
\end{eqnarray}
Note that change the sign of $\mu$ or $a$ will not effect the
character of the solution, so without loss of generality we choose
$\mu>0$ and $a>0$ in this paper. The shape of the warp factor
$e^{2A(y)}$ is shown in Fig.~\ref{fig:A_y}.

In order to localize fermions on the thick brane, we need to
introduce the coupling of fermions and background scalar fields. In
despite of the expression of the potential $V_L$, the zero mode
solution for left fermions can be solved formally as:
\begin{eqnarray}
 \tilde{f}_{L0}(z)\propto \exp\left(-\eta\int^z_0 dz' e^{A(z')}
F(\chi(z'),\phi(z'))\right).\label{eq:zeroMode1}
\end{eqnarray}
In order to know whether the zero mode is localized on the brane, we
need to consider the normalizable problem of the solution. The
normalization condition for the zero mode (\ref{eq:zeroMode1}) is
\begin{eqnarray}
\int_{-\infty}^{\infty} dz \exp\left(-2\eta\int^z_0 dz'
e^{A(z')}F(\chi(z'),\phi(z'))\right)<\infty.\label{eq:Norconditon-z}
\end{eqnarray}
Because we have not the analytic expressions of the functions
$A(z)$, $\phi(z)$ and  $\chi(z)$ in $z$ coordinate, we have to
deal with the problem in $y$ coordinate, in which the condition
(\ref{eq:Norconditon-z}) becomes
\begin{eqnarray}
\int_{-\infty}^{\infty} dy \exp\left( -A(y)-2\eta\int^y_0 dy'
F(\chi(y'),\phi(y'))\right)<\infty.\label{eq:Norconditon-y}
\end{eqnarray}
Now, it is clear that, because $A(y)$ does not vanish at
$y\rightarrow\infty$, we must introduce some kind of scalar-fermion
coupling in order to localized fermion zero mode on the brane. The
normalization of the zero mode is decided by the asymptotic
characteristic of $F(\chi(y),\phi(y))$ at $y\rightarrow\pm\infty$. In
this paper, we consider the simplest couplings
$\eta\bar{\Psi}\chi\phi\Psi$ and
$\eta\bar{\Psi}\chi\Psi+\eta'\bar{\Psi}\phi\Psi$.

\subsection {Type I: $\eta\bar{\Psi}\chi\phi\Psi$}

\begin{figure}
\begin{center}
\includegraphics[width=0.45\textwidth]{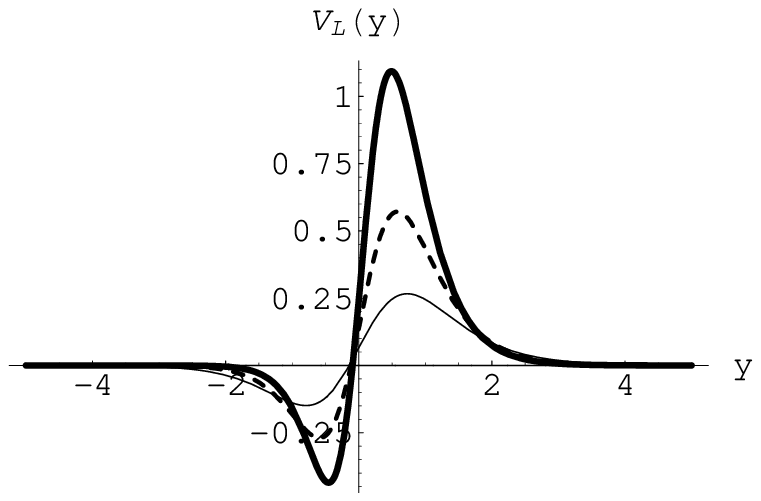}
\includegraphics[width=0.45\textwidth]{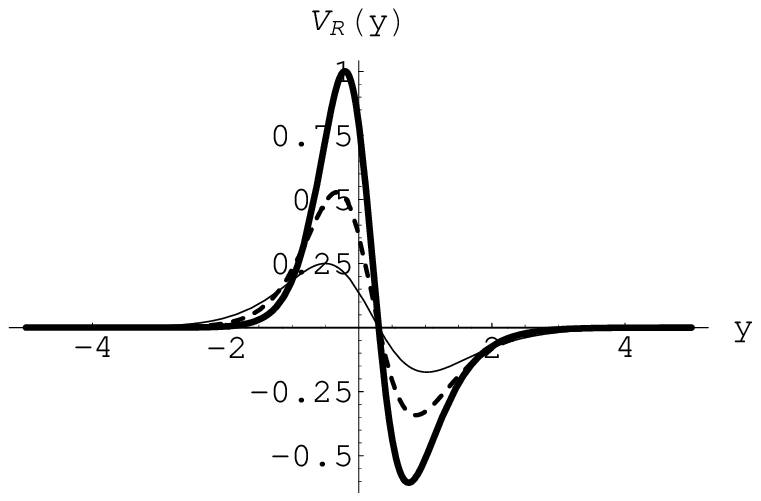}
\end{center}
\caption{  The shape of the potentials $V_L(y)$ and $V_R(y)$ for the
coupling $F(\chi,\phi)=\chi\phi$. The parameters are set to $\mu=1$,
$\eta=2$, $a=1.2$ (thick line), $a=1$ (dashed line) and $a=0.8$ (thin
line).} \label{fig:VCase2}
\end{figure}

This type of coupling with $F(\chi,\phi)=\chi\phi$ has been
considered for example in \cite{Almeida200979}, which is a natural
generation of one-scalar brane scenarios. The explicit forms of the
potentials (\ref{eq:V}) are
\begin{eqnarray}
V_L&=&\frac{a^3 \eta}{144} ~
   e^{-\frac{1}{9} a^2\tanh^2(a{\mu}y)}
   [1+\tanh(a{\mu}y)]^{4a^2/9}  \nonumber\\
   &\times& \biggl\{a ~ \textrm{sech}^2(a{\mu}y)
   \big[9\eta+4a\mu\tanh(a{\mu}y)\big] \nonumber\\
    &-&  8a^2 \mu +8\mu(9+a^2)\tanh(a{\mu}y)
   \biggl\}\textrm{sech}^2(a{\mu}y) ,~~\\
V_R &=& V_L|_{\eta\rightarrow-\eta}.
\end{eqnarray}
The shape of the potentials $V_L$ and $V_R$ is shown in Fig.
\ref{fig:VCase2}. Although there exist potential wells, the zero
modes for both left and right fermions can not be localized on the
brane because the potentials trend to zero from below at
$y\rightarrow\infty$ or $y\rightarrow-\infty$. This can also be
validated by the following expression of the integrand in
(\ref{eq:Norconditon-y}) for the left-handed fermion zero mode:
\begin{eqnarray}
I_0&\equiv&\exp\left[-A(y)-2\eta\int^y_0 dy'
\chi(y')\phi(y')\right]\nonumber\\
&=&\exp\bigg[
     \frac{a^2}{18} \tanh^2(a{\mu}y)
     {-\frac{a\eta}{2\mu} \tanh(a{\mu}y)} 
   -\frac{2a^2}{9}\ln  (1+\tanh(a{\mu}y))
     \bigg]\nonumber\\
   &\rightarrow& \exp\left(-{4}a^3 {\mu}y/{9} \right)
    \rightarrow \infty ~~ \textrm{when} ~~ y \rightarrow -\infty,
\end{eqnarray}
which results in that the normalization condition
(\ref{eq:Norconditon-y}) for the zero mode can not be satisfied.

\subsection {Type II: $\eta\bar{\Psi}\chi\Psi+\eta'\bar{\Psi}\phi\Psi$
}

Since both $\phi$ and $\chi$ are kink solutions, it is natural to ask
whether the kink-fermion coupling of the fermion with each scalar
$\eta\bar\Psi\chi\Psi+\eta'\bar\Psi\phi\Psi$ could trap the fermion
zero mode on the asymmetric brane. We rewrite the coupling as
$\eta\bar\Psi(\chi+\tau\phi)\Psi$ with $\tau=\eta'/\eta$ for
$\eta\neq 0$. The range of the parameter $\tau$ is $-\infty
<\tau<\infty$. While, considering that the potentials of left- and
right-handed fermion KK modes are partner potentials, the range of
$\eta$ can be restricted as $\eta>0$.

For the coupling, the potentials (\ref{eq:V}) are
\begin{eqnarray}
 V_L&=& \frac{{\eta}a^2}{36}
        e^{-\frac{a^2}{9}{\tanh}^2(a{\mu}y)}
        \big[1+{\tanh}(a{\mu}y)]^{\frac{4a^2}{9}} 
        \bigg\{18 \mu (\tau-1){\textrm{sech}}^2(a{\mu}y)\nonumber \\
    && +9{\eta}\big[(\tau+1)-(\tau-1) {\tanh}(a{\mu}y)\big]^2\nonumber\\
    && +2 a^2 \mu\big[(\tau+1)-(\tau-1) {\tanh}(a{\mu}y)\big]\nonumber\\
    && \times\left[\left(2+{\textrm{sech}}^2(a{\mu}y)\right)
    {\tanh}(a{\mu}y)-2\right]\bigg\}, \\
 V_R&=&V_L|_{\eta\rightarrow-\eta}.
\end{eqnarray}
For simplicity, we first focus on the special cases of $\tau=1$ and
$\tau=-1$. Then, we discuss the general case of $\tau$.

\textbf{Case 1: {$\tau=1$}}

For the case $\tau=1$, i.e., $F(\chi,\phi)=\chi+\phi$, we have
$\eta\bar\Psi F(\chi,\phi)\Psi=a\eta\bar\Psi\Psi$, which is nothing
but a mass term of fermions, and the integrand in
(\ref{eq:Norconditon-y}) can be expressed explicitly as
\begin{eqnarray}
I_{1} 
=\exp\bigg\{\frac{a^2}{18}\big[
  \tanh^2(a{\mu}y)
  - 4 \ln (1+\tanh(a{\mu}y))\big] {-2 a\eta y}\bigg\}.
  \nonumber
\end{eqnarray}
Since $I_1>0$ for all $y$, the normalization condition
(\ref{eq:Norconditon-y}) can be decomposed as the following two:
\begin{eqnarray}
 \int_{0}^{\infty} dy~ I_1 <\infty, ~~
 \int_{-\infty}^{0} dy ~I_1 <\infty. \label{eq:Norconditon-yI1}
\end{eqnarray}
When $y\rightarrow+\infty$, we have $I_1\rightarrow \exp(-2 a\eta
y).$ So, in order to ensure that the first condition in
(\ref{eq:Norconditon-yI1}) is satisfied, we require $\eta>0$ for
$a>0$. In the case of $\eta>0$, $a>0$ and $\mu>0$, we reach
\begin{eqnarray}
  I_1\rightarrow \exp\left[-2a\left(\frac{2}{9}a^2 \mu
     +\eta\right) y \right] \rightarrow \infty
\end{eqnarray}
when $y\rightarrow-\infty$, which results in the divergence of the second
integrate in (\ref{eq:Norconditon-yI1}).
Hence, the zero mode of left-handed fermions can not be localized on
the brane, so does the right-handed zero mode.

\textbf{Case 2: {$\tau=-1$}}

Considering the configuration of $\phi$ and $\chi$, we can construct
a combined kink configuration via the two scalars and introduce the
real kink-fermion coupling in the asymmetric braneworld model. The
construction is turned out to be $F(\chi,\phi)=\chi-\phi=a
\tanh(a{\mu}y)$, which is just the case $\tau=-1$ mentioned above,
and usually results in localized left-handed fermion zero mode for
some positive coupling constant $\eta$.

\begin{figure}
\begin{center}
\includegraphics[width=0.45\textwidth]{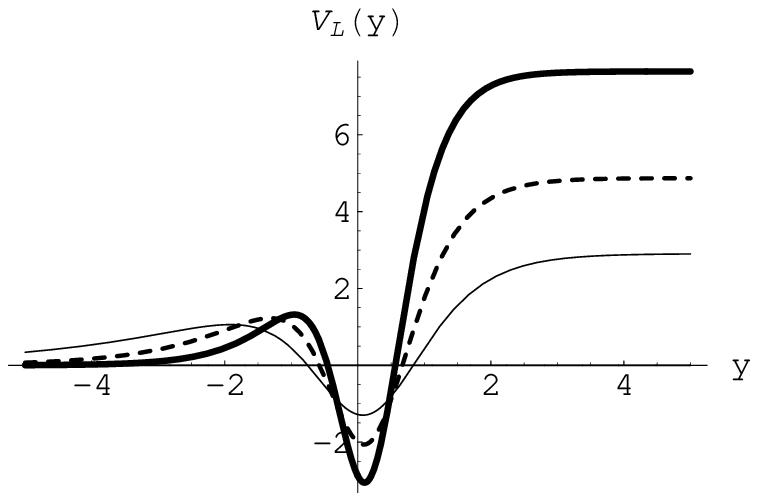}
\includegraphics[width=0.45\textwidth]{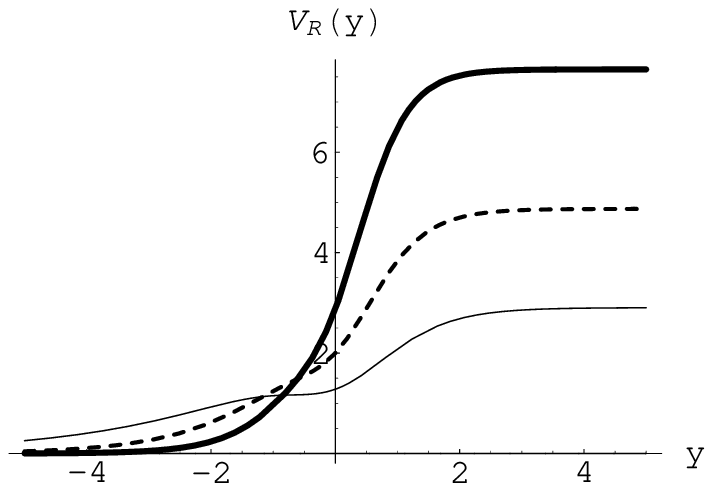}
\end{center}
\caption{ \label{fig:VCase3} The shape of the potential $V_L(y)$ and
$V_R(y)$ for the coupling $F(\chi,\phi)=\chi-\phi$. The parameters
are set to $\mu=1$, $\eta=2$, $a=1.2$ (thick line), $a=1$ (dashed
line) and $a=0.8$ (thin line).}
\end{figure}

\begin{figure}
\begin{center}
\includegraphics{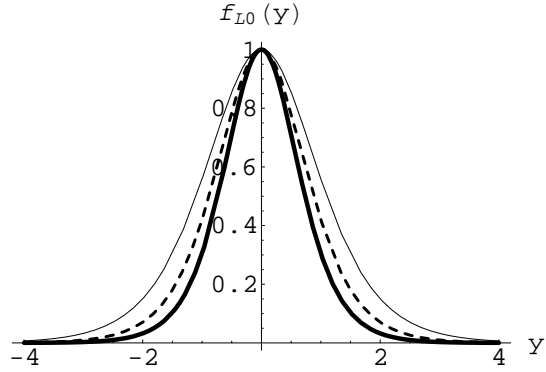}
\end{center}
\caption{ \label{fig:WaveFunctionCaseIIIzeroMode} The zero mode
$f_{L0}$ of left-handed fermions for the  coupling
$F(\chi,\phi)=\chi-\phi$. The parameters are set to $\mu=1$,
$\eta=2$, $a=1.2$ (thick line), $a=1$ (dashed line) and $a=0.8$ (thin
line).}
\end{figure}

The shape of the corresponding potentials is shown in
Fig.~\ref{fig:VCase3}. It can be seen that the potential for
left-handed fermions has a negative potential well and vanishes from
above at $y\rightarrow-\infty$ and trends to a positive constant at
$y\rightarrow\infty$. To the best of our knowledge, such a potential
has not appeared in previous studies. It supports a bound state--the
massless KK mode and continuous massive KK modes with $m^2>0$. For
$m^2\leq V_L(y=\infty)$, the KK modes would be damped at $y>0$ and
oscillated at $y<0$. For $m^2> V_L(y=\infty)$, the KK modes oscillate
at both sides of the brane. Now, we work out the localization
condition for the left-handed fermion zero mode
\begin{eqnarray}
\tilde{f}_{L0}(z(y))&\propto& \exp\left(-\eta\int^y_0 dy'
\big[\chi(y')-\phi(y')\big]\right) \nonumber \\
 &=& \exp\left(-\frac{\eta}{\mu }\ln\cosh (a{\mu}y)\right)\nonumber \\
 &=& \cosh^{-{\eta}/{\mu }} (a{\mu}y).
\end{eqnarray}
The integrand in (\ref{eq:Norconditon-y}) is expressed as
\begin{eqnarray}
I_2 =\exp\bigg\{\frac{a^2}{18}
   \bigg[\tanh^2(a{\mu}y)- 4 \ln  (1+\tanh(a{\mu}y)) \bigg]
   -\frac{2 \eta}{\mu}\ln\cosh (a{\mu}y)\bigg\},
\end{eqnarray}
from which we have
\begin{eqnarray}
 I_2&\rightarrow& \exp(-2 a\eta y)
      ~~~~~~~~~~~~~~~~~\textrm{when} ~ y\rightarrow+\infty, \\
 I_2&\rightarrow& \exp\left[
      2a\left(\eta-\frac{2}{9}{\mu}a^2\right) y\right]
      ~~~\textrm{when} ~ y\rightarrow-\infty .
\end{eqnarray}
So if
\begin{eqnarray}
 \eta>2{\mu}a^2/9 , \label{normalizeConditionTau-1}
\end{eqnarray}
the zero mode of left-handed fermions is normalizable. While the
right one can not be localized, which is a simple consequence of the
fact that $V_L$ and $V_R$ are partner potentials.
Fig.~\ref{fig:WaveFunctionCaseIIIzeroMode} clearly shows that the
zero mode of left-handed fermions is localized on the brane.

\textbf{Case 3: arbitrary $\tau$}

Now, we investigate the the general case with arbitrary $\tau$. The
left-handed fermion zero mode is read as
\begin{eqnarray}
\tilde{f}_{L0}(y)&\propto& \exp\left(-\eta\int^y_0 dy'
\big[\chi(y')-\tau\phi(y')\big]\right) \nonumber \\
 &=& e^{-(\tau+1)a{\eta}y/2}
   \cosh^{(\tau-1){\eta}/(2{\mu})} (a{\mu}y).
\end{eqnarray}
The integrand in (\ref{eq:Norconditon-y}) is expressed as
\begin{eqnarray}
I_3 &=& \frac{a^2}{18}\tanh^2 (a\mu y)
       -\frac{a}{9} \left[9 (\tau+1) \eta
       +2 a^2 \mu \right]y \nonumber\\
   & &
       +\left[\frac{(\tau-1)\eta}{\mu} +\frac{2a^2 \mu}{9} \right]
        \ln(\cosh(a\mu y)).
\end{eqnarray}
Now we have
\begin{eqnarray}
 I_3&\rightarrow& \exp(-2 a\eta y)
      ~~~~~~~~~~~~~~~~~~~\textrm{when} ~ y\rightarrow+\infty, \\
 I_3&\rightarrow& \exp\left[
      2a\left(-\tau\eta-\frac{2}{9}{\mu}a^2\right) y\right]
      ~\textrm{when} ~ y\rightarrow-\infty .
\end{eqnarray}
So if
\begin{eqnarray}
 \eta>0~~\textrm{and} ~~\eta'=\tau\eta<-2{\mu}a^2/9, \label{normalizeConditionForTau}
\end{eqnarray}
the zero mode of left-handed fermions is normalizable. For the case
$\tau=1$ and $\eta>0$, the condition (\ref{normalizeConditionForTau})
can not be satisfied, which results in that the massless mode can not
be localized on the brane. For the case $\tau=-1$, the normalization
condition (\ref{normalizeConditionForTau}) reduces to
(\ref{normalizeConditionTau-1}).

\section{Conclusions}\label{section:conclusion}

In this paper we have investigated the problem of localization of
fermions on the asymmetric two-field thick branes. Different from
other two-scalar brane models, in the asymmetric branes, both scalars
are kink-like configurations and there are two ``Minkowski-type"
regions separated by a transition one along extra dimension. These
characters make it difficult to localize fermions on the branes,
because both the kink-fermion coupling $\bar\Psi(\phi+\chi)\Psi$ and
usual coupling $\bar\Psi\phi\chi\Psi$ do not support the localization
of 4-dimensional fermions on the branes. The key point of
localization problem for fermions is to construct effective couplings
with the two scalars. By analysis, we showed that the real
kink-fermion coupling corresponding to one-scalar-generated brane
scenarios is $\eta\bar{\Psi}(\chi-\phi)\Psi$, for which the potential
for the KK modes of left-handed fermions has a negative well and
there exists only a bound massless mode followed by a continuous
gapless spectrum of KK states. The zero mode of left-handed fermions
is normalizable and can be trapped on the brane under the condition
$\eta>\eta_0$ with the critical coupling constant $\eta_0=2
{\mu}a^2/9 $. For general coupling
$\eta\bar\Psi\chi\Psi+\eta'\bar\Psi\phi\Psi$, the normalization
condition for the massless left-handed fermion KK mode is $\eta>0$, $
\eta'<-\eta_0$.

\section*{Acknowledgements}

This work was supported by the Program for New Century Excellent
Talents in University, the National Natural Science Foundation of
China (No. 10705013), the Doctoral Program Foundation of Institutions
of Higher Education of China (No. 20070730055 and No.
20090211110028), the Key Project of Chinese Ministry of Education
(No. 109153), the Natural Science Foundation of Gansu Province, China
(No. 096RJZA055), and the Fundamental Research Funds for the Central
Universities (No. lzujbky-2009-54).


\section*{References}

\begin{thebibliography}{10}

\bibitem{ADD}
 Arkani-Hamed H, Dimopoulos S and Dvali G 1998
    {\it Phys. Lett. B} \textbf{429} 263;
 Antoniadis I, Arkani-Hamed N, Dimopoulos S and Dvali G 1998
    {\it Phys. Lett. B} \textbf{436} 257.


\bibitem{RS}
 Randall L and Sundrum R 1999
    {\it Phys. Rev. Lett.} \textbf{ 83} 3370;
    1999 {\it Phys. Rev. Lett.} \textbf{ 83} 4690.

\bibitem{Rubakov1983}
 Rubakov V A and Shaposhnikov M E 1983
    {\it Phys. Lett. B} \textbf{125} 136;
    1983 {\it Phys. Lett. B} \textbf{125} 139.


\bibitem{DeWolfe200062}
 DeWolfe O, Freedman D Z, Gubser S S and Karch A 2000
    {\it Phys. Rev. D} \textbf{62} 046008.

\bibitem{Csaki2000581}
 Csaki C, Erlich J, T. Hollowood and Shirman Y 2000
    {\it Nucl. Phys. B} \textbf{581} 309.

\bibitem{Gremm2000478}
 Gremm M 2000
    {\it Phys. Lett. B} \textbf{478} 434.


\bibitem{Campos200288}
 Campos A 2002
    {\it Phys. Rev. Lett.} \textbf{88} 141602.


\bibitem{ThickBranes}
 Melfo A, Pantoja N and Skirzewski A 2003
    {\it Phys. Rev. D} \textbf{67} 105003;
 Bronnikov K A and Meierovich B E 2003
    {\it Grav. Cosmol.} \textbf{9} 313;
 Bazeia D, Furtado C and Gomes A R 2004
    {\it JCAP} \textbf{0402} 002;
 Bazeia D, Gomes A R, Losano L and Menezes R 2009
    {\it Phys. Lett. B} \textbf{671} 402;
 Dzhunushaliev V, Folomeev V, Singleton D and Aguilar-Rudametkin S 2008
    {\it Phys. Rev. D} \textbf{77} 044006;
 Bazeia D, Brito F A and Losano L 2006
    {\it JHEP} \textbf{0611} 064;
 Zhao L, Liu Y -X and Duan Y -S 2008
    {\it Mod. Phys. Lett. A} \textbf{ 23} 1129;
 Sarrazin M and Petit F 2010
     {\it Phys. Rev. D} \textbf{81} 035014;
 Shtanov Y, Sahni V, Shafieloo A and Toporensky A 2009
    {\it JCAP} \textbf{04} 023;
 Cruz W T, Gomes A R and Almeida C A S  2009
    Resonances on deformed thick branes
    arXiv:0912.4021[hep-th];
 Cruz W T, Tahim M O and Almeida C A S 2009
    {\it Europhys. Lett.} \textbf{88} 41001.

\bibitem{ThickBraneWeyl}
 Arias O, Cardenas R and Quiros I 2002
    {\it Nucl. Phys. B} \textbf{643} 187;
 Barbosa-Cendejas N and Herrera-Aguilar A 2005
    {\it JHEP} \textbf{0510} 101;
 Barbosa-Cendejas N and Herrera-Aguilar A 2006
    {\it Phys. Rev. D} \textbf{73} 084022.

\bibitem{Liu0907.1952}
 Liu Y -X, Zhong Y and Yang K 2009
    Scalar-Kinetic Branes
    arXiv:0907.1952[hep-th].

\bibitem{Dzhunushaliev:09a}
 Dzhunushaliev V, Folomeev V and Minamitsuji M 2010
    {\it Rept. Prog. Phys.} \textbf{73} 066901.



\bibitem{Bazeia200405}
 Bazeia D and Gomes A R 2004
    {\it JHEP} \textbf{05} 012.

\bibitem{Blochwall}
  Bazeia D, Boschi-Filho H and Brito F A 1999
    {\it JHEP} \textbf{ 9904} 028;
  Bazeia D, Nascimento J R S, Ribeiro R F and Toledo D 1997
    {\it J. Phys. A}  \textbf{ 30} 8157;
  Bazeia D, dos Santos M J and Ribeiro R F 1995
    {\it Phys. Lett.  A} \textbf{208} 84.


%

\bibitem{SouzaDutra200878}
 de Souza Dutra A, Amaro de Faria Jr. A C and Hott M 2008
    {\it Phys. Rev. D} \textbf{78} 043526;
 de Souza Dutra A, Amaro de Faria Jr. A C and Hott M 2009
    Degenerate and critical domain walls and accelerating
    universes driven by bulk particles
    arXiv:0903.5533[hep-th].

\bibitem{LocalizedFermion}
  Randjbar-Daemi S and Shaposhnikov M 2000
     {\it Phys. Lett. B} \textbf{492} 361;
  Ringeval C, Peter C P and Uzan J P 2002
    {\it Phys. Rev. D} \textbf{65} 044016;
  Koley R and Kar S 2005
    {\it Class. Quantum Grav.} \textbf{22} 753;
  Melfo A, Pantoja N and Tempo J D 2006
    {\it Phys. Rev. D} \textbf{73} 044033;
  Slatyer T R and Volkas R R 2007
    {\it JHEP} \textbf{0704} 062;
  Zhang X -H, Liu Y -X and Duan Y -S 2008
    {\it Mod. Phys. Lett. A} \textbf{23} 2093;
  Koley R, Mitra J, and SenGupta S 2008
    {\it Phys. Rev. D} \textbf{78} 045005;
  Davies R and George D P 2007
    {\it Phys. Rev. D} \textbf{76} 104010;
 Liu Y -X, Yang J, Zhao Z -H, Fu C -E and Duan Y -S 2009
    {\it Phys. Rev. D} \textbf{80} 065019;
 Zhang L -J and Yang G H 2009
    Zero Modes of Matter Fields on Scalar Flat Thick Branes
    arXiv:0907.1178[hep-th];
 Liang J, Duan Y -S 2009
    {\it Phys. Lett. B} \textbf{ 680} 489.

\bibitem{Liu2008D78}
 Liu Y -X, Zhang L -D, Zhang L -J and Duan Y -S 2008
    {\it Phys. Rev. D} \textbf{78} 065025;
 Liu Y -X, Zhang X -H, Zhang L -D and Duan Y -S 2008
    {\it JHEP} \textbf{0802} 067.



\bibitem{Almeida200979}
 Almeida C A S, Casana R, Ferreira Jr. M M and Gomes A R 2009
    {\it Phys. Rev. D} \textbf{79} 125022.

\bibitem{LiuJHEP2009}
 Liu Y -X, Li H -T, Zhao Z -H, Li J -X and Ren J R 2009
    {\it JHEP} \textbf{0910} 091.


\bibitem{Parameswaran0608074}
 Parameswaran S L, Randjbar-Daemi S and Salvio A 2007
    {\it Nucl. Phys. B} \textbf{767} 54.

\bibitem{GeorgePRD2007}
 George D P and Volkas R R 2007
    {\it Phys. Rev. D} \textbf{75} 105007.

\bibitem{Liu200808}
 Liu Y -X, Zhang L -D, Wei S W and Duan Y -S 2008
    {\it JHEP} \textbf{0808} 041.

\bibitem{Liu20090902}
 Liu Y -X, Zhao Z -H, Wei S -W and Duan Y -S 2009
    {\it JCAP} \textbf{0902} 003.

\bibitem{KodamaPRD2009}
 Kodama Y, Kokubu K and Sawado N 2009
    {\it Phys. Rev. D} \textbf{79} 065024.


\bibitem{Ringeval200265}
 Ringeval C, Peter P and Uzan J P 2002
    {\it Phys. Rev. D} \textbf{65} 044016.



\bibitem{Liu0907.4424}
 Liu Y -X, Guo H, Fu C -E and Ren J R 2010
    {\it JHEP} \textbf{02} 080;
 Liu Y -X, Fu C -E, Zhao L and Duan Y -S 2009
    {\it Phys. Rev. D} \textbf{80} 065020.

\end{thebibliography}
\end{document}